# Stacking order and interlayer coupling tuning the properties of charge density waves in layered 1*T*-NbSe$_2$

Tao Jiang[1], Haotian Wang[1], Heng Gao[1], Qinghe Zheng[3], Zhenya Li[1] and Wei Ren[1,2,*]

[1] *Physics Department, International Center for Quantum and Molecular Structures, Shanghai Key Laboratory of High Temperature Superconductors, Shanghai University, Shanghai 200444, China*

[2] *State Key Laboratory of Advanced Special Steel, Materials Genome Institute, Shanghai University, Shanghai 200444, China*

[3] *School of Materials Science and Engineering, Shanghai University, Shanghai 200444, China*

\* Email: renwei@shu.edu.cn

## ABSTRACT

Layered transition metal dichalcogenide 1*T*-NbSe$_2$ is a good candidate to explore the charge density wave (CDW) and Mott physics. However, the effects of stacking orders and interlayer coupling in CDW 1*T*-NbSe$_2$ are still less explored and understood. Using density functional theory calculations, we present a systematic study of the electronic and magnetic properties of monolayer and layered CDW 1*T*-NbSe$_2$. Our results indicate that monolayer CDW 1*T*-NbSe$_2$ is a magnetic insulator with $\sqrt{13} \times \sqrt{13}$ periodic lattice modulation. Nevertheless, the magnetic properties of bilayer CDWs 1*T*-NbSe$_2$ are found stacking orders dependence. The mechanism is understood by the changes of local magnetic moments in each layer due to spin charge transfer between interlayers. Furthermore, the bulk CDW 1*T*-NbSe$_2$ opens a band gap with 0.02 eV in 1×1×2 supercell due to the interlayer spin coupling. We also discover that the electronic structures of layered 1*T*-NbSe$_2$ show a strong dependence on stacking configurations and dimensionality.

## I. INTRODUCTION

During the past decades, the charge density wave (CDW) in transition-metal dichalcogenides (TMDs) including TaS$_2$ and TaSe$_2$ have been extensively studied involving Mott insulators [1-4], superconductivity [5-8], and promising applications [9-12] in their two-dimensional [9,13-17] and bulk [18-23] cases. For instance, bulk 1*T*-TaS$_2$ undergoes a series of structural transitions including first-order and second-order transitions during temperature declining process. Below 180 K, it forms the commensurate charge density wave (CCDW) characterized by 13 Ta atoms accumulating in a $\sqrt{13} \times \sqrt{13}$ "Star of David" pattern. The periodic lattice modulation leads to the significant change of the electronic structures, resulting in an insulating ground state of 1*T*-TaS$_2$. Due to the half-filled nonbonding state of center Ta atom, such an insulating state is usually ascribed as a Mott insulator [2,24,25]. Besides, the interlayer dimerization and stacking order have been proved to be indispensable to understand the insulating state of CDW in TaS$_2$ [17,19,26]. Besides, the CDW in two-dimension (2D) exhibits unique properties different from the corresponding bulk cases.

For example, 1$T$-TaS$_2$ undergoes a series of metastable states when the thickness is reduced down to nanometer scale [9]. A robust Mott insulating state induced by Coulomb correlation effect was found in monolayer 1$T$-TaSe$_2$ with special orbital texture [4].

However, 1$T$-NbSe$_2$, as a member of CDWs in TMDs, has been paid less attention either in 2D or bulk cases. As a matter of fact, 2$H$-NbSe$_2$ is usually easier to be fabricated than 1$T$ phase, thus the 2$H$ phase has been intensively investigated for its CDWs [27-36]. Bulk 2$H$-NbSe$_2$ undergoes the nearly commensurate charge density wave (NCCDW) phase at 33 K with a 3×3 periodic superlattice, which shows the independence of temperature [34]. With the temperature decreasing to 7 K, it hosts the coexistence of superconductivity and incommensurate charge density wave (ICCDW) [25,27]. And the CDW transition temperature was enhanced to 145 K from 33 K when NbSe$_2$ was exfoliated into monolayer [30], indicating the enhancement the critical temperature of CDW order in 2D case. Recently, monolayer 1$T$-NbSe$_2$ film was firstly synthesized on bilayer graphene, and found to be a Mott insulator with a band gap of 0.4 eV [37]. Liu *et al*. [38] reported that the Mott upper Hubbard band of monolayer 1$T$-NbSe$_2$ is distributed away from the center d$z^2$ orbital in the $\sqrt{3} \times \sqrt{3}$ R30° periodicity. It has been proved that different stacking types of CDWs could also have an impact on the metal-insulator transition and insulating phase, accompanied with an interlayer Peierls dimerization [19]. And a possible metal-insulator transition was observed when the interlayer antiferromagnetic order and electronic correlation effects were taken into account in bulk 1$T$-NbS$_2$ [39]. However, less attention has been paid to the stacking effects and interlayer coupling in layered 1$T$-NbSe$_2$.

In this work, we use the first-principles calculations to investigate the monolayer, bilayer, and bulk 1$T$-NbSe$_2$, especially to concentrate on the influence of stacking and interlayer coupling towards the properties of CDW. The monolayer 1$T$-NbSe$_2$ shows metallicity in normal structure but insulator in the CDW phase using both PBE and PBE+$U$ methods. Each Star of David has 1 $\mu_B$ total magnetic moment contributed mainly by the central Nb atoms. Then we construct five different stacking orders bilayer 1$T$-NbSe$_2$ and find that the interlayer spin charger transfer can effectively influence the local magnetic moments of bilayer CDW. In bulk case, the results indicate the CDW phase is an out-of-plane metal without the consideration of interlayer magnetic order, which is same as the first-principles results of bulk 1$T$-TaS$_2$ [13,14,26,40]. Interestingly, when the interlayer antiferromagnetic order is considered, the band structure opens a gap, possibly making the system a Mott insulator. Moreover, by comparing five stackings of bilayer and bulk CCDW, we elucidate the stacking dependence of the electronic structure of 1$T$-NbSe$_2$. Our work shows the mechanisms of interlayer coupling and stacking order modulate the electronic structure and magnetic properties of 1$T$-NbSe$_2$, suggesting the critical role of interlayer coupling and stacking order in understanding the magnetic properties in CDW materials.

## II. COMPUTATIONAL METHODS

The density functional theory (DFT) calculations were performed with the projector augmented wave (PAW) [41] method implemented in the Vienna *Ab initio* Simulation Package (VASP) [42,43]. Within the generalized gradient approximation (GGA) [44], we employed Perdew-Burke-Ernzerhof (PBE) exchange and correlation functional. The energy cutoff of 520 eV was chosen throughout all calculations. For the undistorted structure and $\sqrt{13} \times \sqrt{13}$ CDW phases, $12 \times 12 \times 1$ and $4 \times 4 \times 1$ Γ-centered Monkhorst-Pack *k* meshes were used in geometry optimization and self-consistent calculations. On the other hand, $12 \times 12 \times 8$ and $4 \times 4 \times 8$ *k* meshes were used for bulk in geometry optimization and self-consistent calculations. The structure relaxation was achieved by converging all the forces and energies within 0.01 eV/Å and $10^{-6}$ eV, respectively. For Nb 4*d* orbitals, an effective Hubbard *U* = 2.95 eV [45] was adopted to treat the Coulomb interaction. For bilayer and bulk 1*T*-NbSe$_2$, a van der Waals (vdW) correction [46] was adopted to reproduce the experimental lattice constants. In monolayer and bilayer, we employed a vacuum layer of 18 Å between the periodic images.

### III. TWO-DIMENSIONAL CDW

TABLE I. Lattice parameters, relative energy, band gap, and total magnetic moment of monolayer 1*T*-NbSe$_2$ for normal and CDW phases calculated by using PBE and PBE+*U* methods.

| Methods | Phase | *a* (Å) | Δ*E* (meV) | $E_g$ (eV) | Mag (μ$_B$) |
|---|---|---|---|---|---|
| PBE | 1*T* | 3.48 | 0 | 0 | 0 |
|  | CCDW | 12.60 | -52.78 | 0.05 | 0.92 |
| PBE+*U* | 1*T* | 3.53 | 0 | 0 | 0.19 |
|  | CCDW | 12.72 | -54.76 | 0.35 | 1 |

We begin our discussions from the normal state of monolayer 1*T*-NbSe$_2$ and corresponding $\sqrt{13} \times \sqrt{13}$ CDW phase. The normal state of monolayer 1T-NSe$_2$ crystalizes $D_{3d}$ point group with octahedral coordination of Nb atom, and the Se-Nb-Se atomic planes are arranged by ABC stacking order. TABLE I lists the structural parameters and electronic properties of normal and CDW phases of monolayer 1*T*-NbSe$_2$ which are calculated by PBE and PBE+U methods, respectively. The lattice constants of normal and CDW phases monolayer 1*T*-NbSe$_2$ are 3.48 Å and 12.60 Å which are slightly larger than that of PBE +vdW results (3.45 Å and 12.46 Å). Comparing the energies between normal phase and $\sqrt{13} \times \sqrt{13}$ CDW phase of monolayer 1*T*-NbSe$_2$, we find that the lattice reconstruction in CDW phase lowers the energy by about 54.8 meV per formula unit (f.u.) than normal phase. It indicates that CDW phase is more energetic stable. Then we turn to the electronic structures of monolayer 1*T* and CDW phases as shown in Fig. 1. The PBE band structure of normal phase show metallicity without band splitting and the band structures are slightly changed in the present of spin-orbital coupling (SOC) effect. While the band structure of CDW opens a narrow gap ~0.05 eV with band splitting, in which the SOC lead to a metallic dispersion crossing Fermi level. Considering the electron correlation effect of Nb 4$d$ orbitals, we perform the band-structure calculations using PBE+$U$ ($U$ = 2.95 eV) method for CDW phase. The PBE+$U$ results show an explicit band gap of about 0.35 eV between two flat bands, consistent with the report of STM measurement [37]. Besides, 1 $\mu_B$ total magnetic moment is found in one Star of David, in which the central Nb atom contributes one-half magnetic moment. When SOC is taken into account, the band gap diminishes about 0.1 eV near Fermi surface.

TABLE II. The relative energy $\Delta E$, equilibrium interlayer spacing $d$, distance between central Nb atoms in each layer $d_{Nb-Nb}$ and ground magnetic configurations of NbSe$_2$ for various CDW stacking patterns.

| Stacking orders | $\Delta E$ (meV/Star) | $d$ (Å) | $d_{Nb-Nb}$ (Å) | Magnetic configuration |
|---|---|---|---|---|
| AA | 0.00 | 2.78 | 6.34 | None |
| AB | 39.03 | 2.70 | 7.20 | AFM |
| AC | 21.39 | 2.72 | 8.74 | AFM |
| AL | 30.47 | 2.72 | 8.76 | AFM |
| AM | 43.63 | 2.70 | 7.20 | AFM |

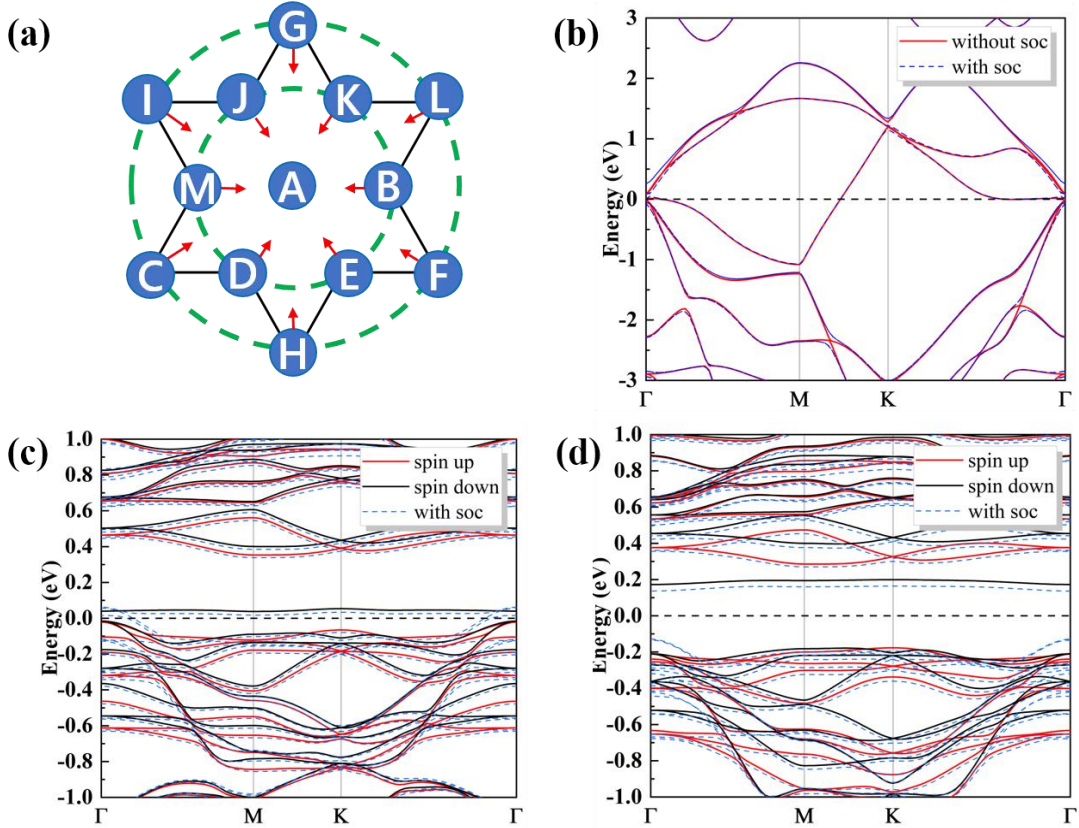

FIG. 1. (a) The 13 sites of Nb atoms are labeled in one Star of David. The PBE band structures of monolayer 1$T$-NbSe$_2$ with (b) normal and (c) CDW phases. (d) The

PBE+$U$ band structure of CCDW phase. The red and black solid lines represent the spin-up and spin-down states respectively. The blue dashed lines indicates the consideration of SOC. The Fermi level is set to 0 eV by the black dashed line.

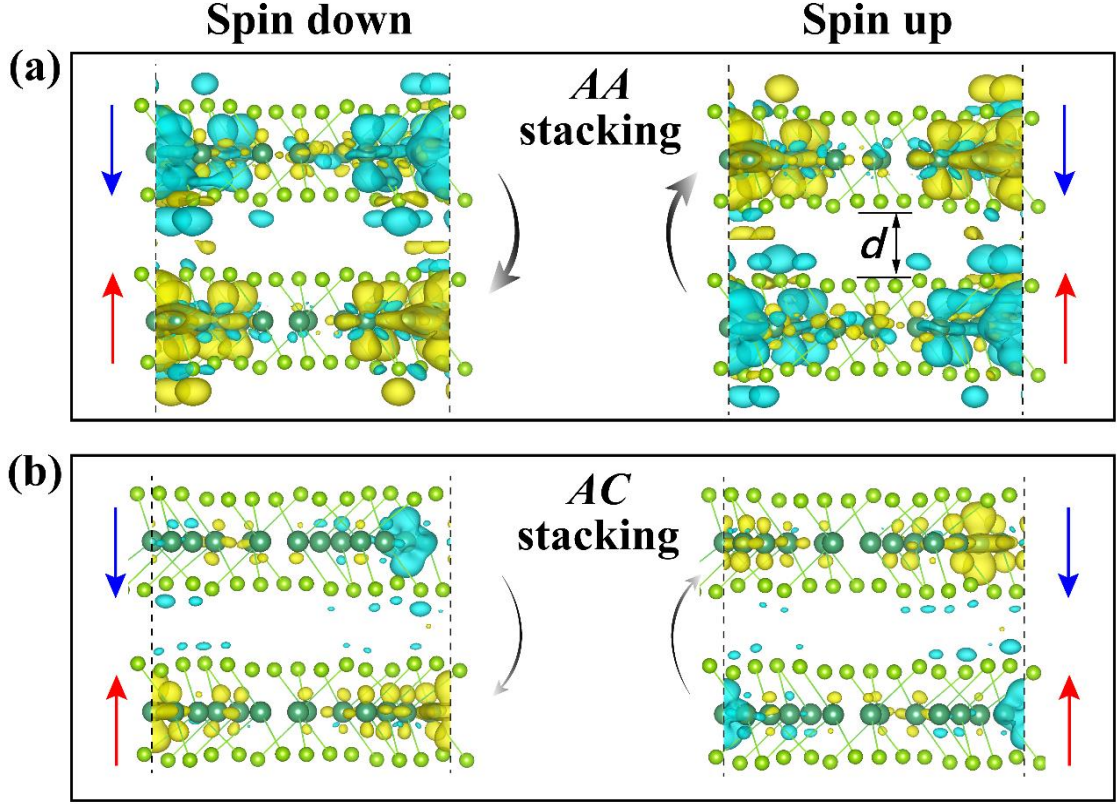

FIG. 2. Spin dependent interlayer differential charge density (SDCD) of NbSe$_2$ bilayer. (a) Spin up and spin down SDCD of AA stacking. (b) Spin up and spin down SDCD of AC stacking. The red and green colors indicate the spin charge accumulation and depletion respectively.

To explore the influence of the stacking orders and interlayer spin coupling in layered CDW systems, we construct bilayer 1$T$-NbSe$_2$ with different high symmetric stacking orders. Taking the central Nb atom in the bottom layer as reference A, the center of Star of David cluster in the top layer can be arranged in 13 sites (Fig. 1(a)), i.e., A, B, …, M. However, there are only five inequivalent stacking orders due to the threefold rotational symmetry, namely AA, AB, AC, AL, and AM stackings. The interlayer van der Waals interaction and initial magnetic configuration are considered in our calculations. As depicted in TABLE II, the relative energy values of 1$T$-NbSe$_2$ bilayers with different stacking orders vary from 0 to ~ 40 meV. The most stable structure is the AA stacking with the largest equilibrium interlayer distance and the smallest $d_{Nb-Nb}$, same as the case of TaS$_2$ and TaSe$_2$ bilayer CDWs [17]. The band structures of five stackings show explicit band gap as shown in Fig. S1 of supplemental materials. However, magnetic property of bilayer NbSe$_2$ differs from TaS$_2$ and TaSe$_2$, in which the magnetic moment disappears in AA stacking but preserves in other four ones with AFM being the ground state. Similarly, the vanish of magnetic moment in AA stacking can also be explained from the view of interlayer spin-dependent differential charge density (SDCD) [17,47]. The SDCD of AA and AC stackings are depicted in Fig. 2 with an isosurface value of 0.0005 e/bohr$^3$, where the top and bottom layers are assigned with spin-down and spin-up orders. In the left panel, we find remarkable spin-down charge transfer from top to bottom layer, in contrast to the case of right panel that spin-up charge transfer from bottom to top layer. The total magnetic moments in top or bottom layer are canceled out because the spin-up and spin-down charge transfer are equal to each other in magnitude. In contrast, few spin charge transfer are found in AC and AL stackings, therefore the decrease of magnetic moments are slighter than the other three stackings. Furthermore, the effects of interlayer charge transfer can be tuned by changing the interlayer distance, as shown in Fig. 3. When the distance is increased, the local magnetic moment of each layer increases and finally approaches the monolayer limit due to the weakening of charge transfer. And the magnetic moment in opposite directions will reduce with the strengthening of charge transfer. Interestingly, the magnetic moment of AC and AL stackings remain even if we reduce the distance by 1 Å and the curves does not display any abrupt change like the other three stackings, which can significantly suggest the stacking order dependence of magnetic properties.

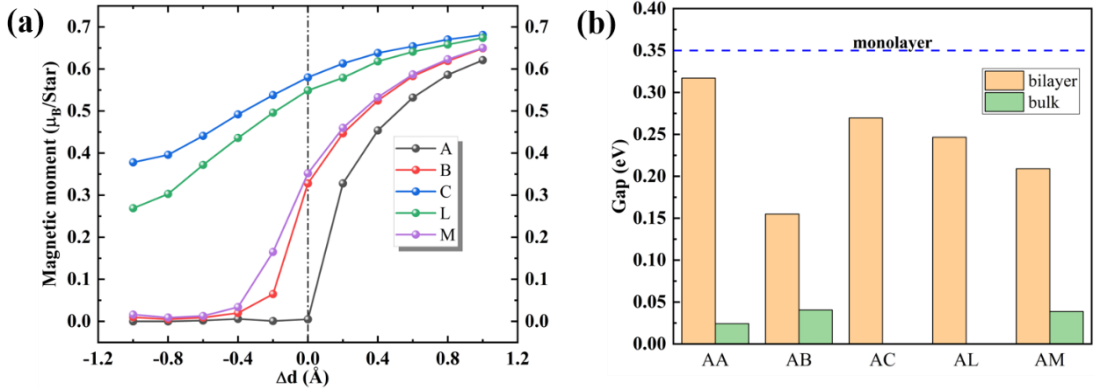

FIG. 3. (a) The local magnetic moment in each CDW layer as a function of the change of interlayer distance. (b) The band gaps of bilayer and bulk NbSe$_2$ with different stacking orders. The blue dashed line represents the gap of monolayer CDW.

## IV. BULK CDW

In this section, we focus on the relationship between CDW properties and stacking orders in bulk 1$T$-NbSe$_2$. The crystal structure and corresponding Brillouin zone are shown in Fig. 4. Bulk 1$T$-NbSe$_2$ has P$\bar{3}$m1 space group in which adjacent layers are combined by weak van der Waals forces. The optimized lattice parameters calculated by PBE +optB86b method are $a$ = 3.45 Å and $c/a$ = 1.81. The spin-polarized electronic structures of normal phase show metallicity without band splitting, similar with the band of monolayer NbSe$_2$. We obtain the CDW of bulk 1$T$-NbSe$_2$ by considering the atomic distortions in $\sqrt{13} \times \sqrt{13}$ supercell normal phase. We found the CDW phase of bulk 1$T$-NbSe$_2$ is in energy lower about 50.61 meV than the normal phase. In addition, the atomic distortion of CDW causes the modification of electronic structures. As shown in Fig. 4(d), the PBE band structure suggests a distinct in-plane gap of about 0.2 eV and an out-of-plane metallic dispersion along Γ-A direction. Such a one-dimensional metallic behavior of bulk 1T-NbSe$_2$ is consistent with the DFT results of TaS$_2$ [13,14,40] and TaSe$_2$ [48]. To revise the self-interaction error caused by Coulomb correlation effects, we perform PBE+$U$ band calculations of CDW phase. We find that the in-plane gap is increased to about 0.3 eV but the out-of-plane metallic band dispersion preserves.

TABLE III. The relative energy Δ$E$, band gap, and magnetic configurations of bulk NbSe$_2$ for various CDW stacking patterns.

| Stacking orders | Δ$E$ (meV/Star) | Gap (eV) | Magnetic configuration |
|---|---|---|---|
| AA | 12.63 | 0.02 | AFM |
| AB | 19.21 | 0.04 | AFM |
| AC | 0.24 | 0.00 | AFM |
| AL | 0.00 | 0.00 | AFM |
| AM | 20.08 | 0.04 | AFM |

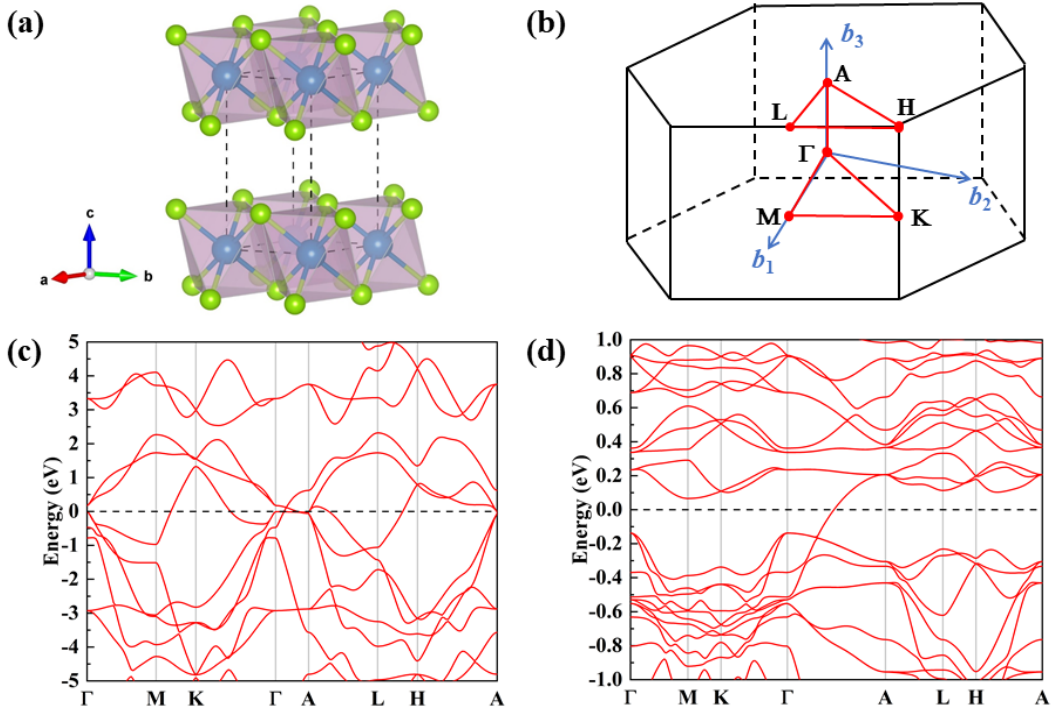

FIG. 4. (a) Crystal structure of bulk 1T-NbSe$_2$ and (b) corresponding Brilouin zone. Band structures of (c) normal and (d) CDW phases calculated by using PBE functional.

In the past few years, the mechanism of insulating behavior of CDW materials like 1T-TaS$_2$ has attracted much interest and debate. Recently, Lee *et al.* [19] reported the Mott gap induced by the interlayer Peierls dimerization and the metal-insulator transition by stacking order in 1T-TaS$_2$. Shin *et al.* [26] used the DFT +ACBN0 method to identify bulk 1T-TaS$_2$ as a Mott insulator with layer-by-layer AFM configuration. Besides, a metal-insulator transition was realized in 1×1×2 supercell of bulk 1T-NbS$_2$ with consideration of interlayer AFM order and electron correlation effect [39]. Based on the findings of CDW materials from both experiments and first-principles, we expect that the stacking order and interlayer coupling play a critical role in the electronic structures and magnetic properties for layered CDWs. Accordingly, we construct the AA stacking 1×1×2 supercell of bulk 1$T$-NbSe$_2$ CDW to explore its electronic structures. As expected, the PBE +$U$ band structure of 1$T$-NbSe$_2$ CDW phase open a gap of about 0.02 eV by considering the interlayer AFM order. The FM configuration is less stable than AFM and still show metallicity. We also consider other four stackings bilayer 1T-NbSe$_2$ CDWs to investigate the effects of stacking orders on electronic structures. In TABLE III, the AL stacking is found to be the most stable among the five stackings, consistent with the results in Ref. [19]. The second stable stacking is AC stacking with a small energy difference 0.24 meV/Star of AL stacking. Nevertheless, the exceptionally stable two stackings are metal and the other three stackings all show small gaps, as shown in Fig. S2. Furthermore, all the five stacking orders have the interlayer AFM ground state.

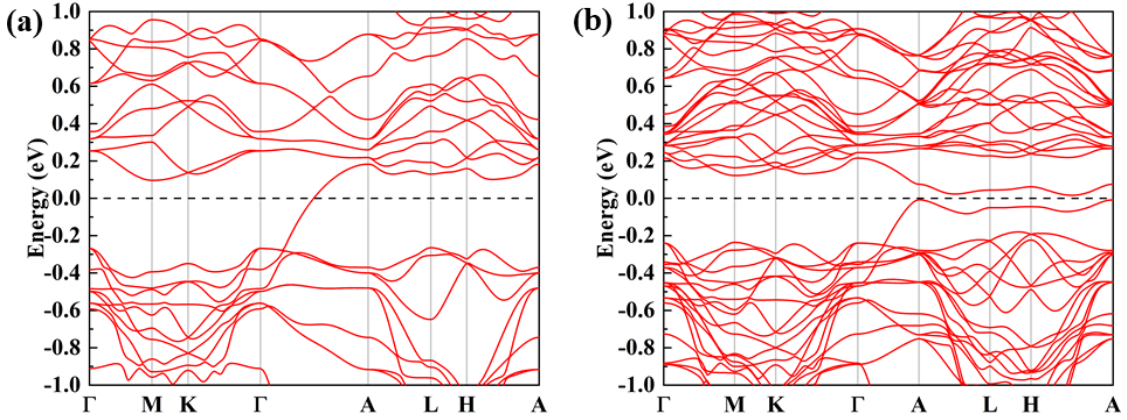

FIG. 5. PBE+$U$ electronic structures of (a) CDW phase and (b) AFM 1×1×2 supercell with AA stacking. The Fermi level is set to 0 eV with black dashed lines.

To provide a comprehensive vision of NbSe$_2$ CDW from the aspects of stacking and dimensionality, we plot the variation of band gaps of monolayer, bilayer and bulk 1$T$-NbSe$_2$ with different stacking orders and present them in Fig. 3(b). In general, the band gap diminishes when the dimensionality increases but shows strong stacking dependence in the same dimensionality. Such a variation of band gap is consistent with other typical CDW system such as few-layers 1$T$-TaSe$_2$ which presents the gap narrowing from monolayer to trilayer in AC stacking.

## V. CONCLUSIONS

In summary, by using first-principles calculations, we studied the electronic and magnetic properties of monolayer, bilayer, and bulk CDW of 1$T$-NbSe$_2$. The monolayer 1$T$-NbSe$_2$ CDW is a magnetic insulator with 1 $\mu_B$ magnetic moment per Star of David. While in bilayer 1$T$-NbSe$_2$ CDW, the magnetic moments in each layer diminish with different stacking orders, which can be explained by the interlayer spin charge transfer.

Then we explored the influence of staking and interlayer interaction towards bulk CDW. The calculations from DFT and DFT +$U$ methods indicate that $\sqrt{13} \times \sqrt{13} \times 1$ bulk CDW is a metal without any magnetic moments. However, the 1×1×2 AA stacking of 1$T$-NbSe$_2$ CDW opens a small gap in the present of interlayer AFM configuration and electron correlation effect. In addition, the AB and AM stacking of bulk 1$T$-NbSe$_2$ CDW show explicit out-of-plane gap as well, in contrast with the metallicity for AC and AL stackings. Among the five stackings of bulk 1$T$-NbSe$_2$ CDW, the most stable is AL staking, consistent with the case of 1$T$-TaS$_2$. By analyzing the variation of band gap from monolayer to bilayer and bulk 1$T$-NbSe$_2$ CDW with various stacking orders, we find that both stacking order and dimensionality can affect the electronic structures of layered 1$T$-NbSe$_2$. Our theoretical work illustrates the significance of interlayer coupling and stacking effects in understanding the properties of 1$T$-NbSe$_2$ CDW systems and we call for more experimental studies of the layered 1$T$-NbSe$_2$ CDW.

## ACKOLEDGEMENTS

This work was supported by the National Natural Science Foundation of China (Grant Nos. 12074241, 52130204, 11929401, 51861145315), the Science and Technology Commission of Shanghai Municipality (Grant Nos. 19010500500, and 20501130600), Shanghai Supercomputer Center, High Performance Computing Center at Shanghai University, and the Key Research Project of Zhejiang Laboratory (No. 2021PE0AC02).

# Supplemental Materials for
# Stacking order and interlayer coupling tuning the properties of charge density waves in layered 1*T*-NbSe$_2$

Tao Jiang[1], Haotian Wang[1], Heng Gao[1], Qinghe Zheng[3], Zhenya Li[1] and Wei Ren[1,2,*]

[1] *Physics Department, International Center for Quantum and Molecular Structures, Shanghai Key Laboratory of High Temperature Superconductors, Shanghai University, Shanghai 200444, China*

[2] *State Key Laboratory of Advanced Special Steel, Materials Genome Institute, Shanghai University, Shanghai 200444, China*

[3] *School of Materials Science and Engineering, Shanghai University, Shanghai 200444, China*

* Email: renwei@shu.edu.cn

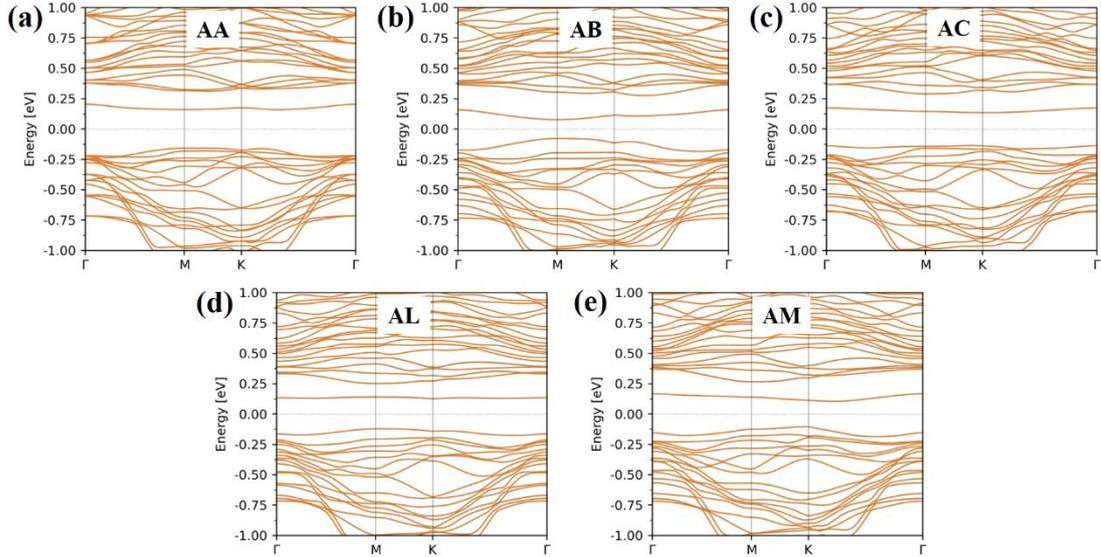

FIG. S1. PBE+*U* band structures of bilayer 1*T*-NbSe$_2$ with five stacking orders. The Fermi energy is set to 0 eV represented by black dotted lines.

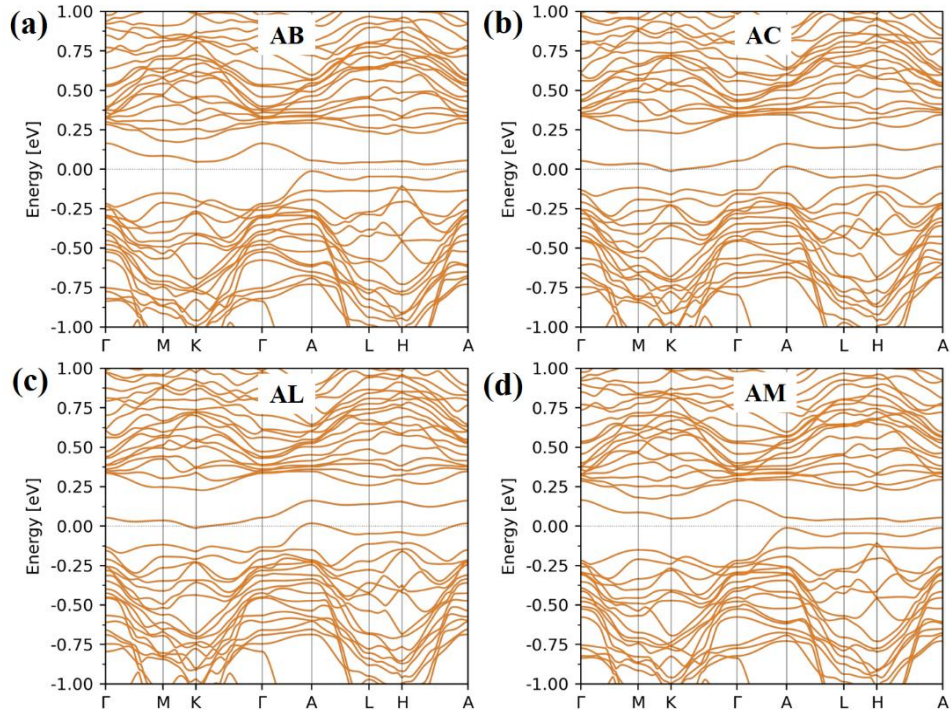

FIG. S2. PBE+$U$ band structures of bulk $1T$-NbSe$_2$ with four stacking orders. The Fermi energy is set to 0 eV represented by black dotted lines.